\documentclass[12pt]{article}             
\usepackage{osajnl}                       
\usepackage{hyperref}
\usepackage{graphicx}

\begin{document}

\title{Atomic Qubit Manipulations with an Electro-Optic Modulator}
\author{P. J. Lee, B. B. Blinov, K. Brickman, L. Deslauriers, M. J. Madsen, R. Miller, D. L. Moehring, D. Stick, and C. Monroe}
\affiliation{FOCUS Center and Department of Physics, University of Michigan, Ann Arbor, MI  48109-1120}

\begin{abstract}
We report new techniques for driving high-fidelity stimulated Raman transitions in trapped ion qubits.  
An electro-optic modulator induces sidebands on an optical source, and interference between the 
sidebands allows coherent Rabi transitions to be efficiently driven between hyperfine ground states 
separated by $14.53\ GHz$ in a single trapped $^{111}Cd^+$ ion.
\end{abstract}


\maketitle

A collection of trapped atomic ions is one of the most attractive candidates for a large-scale quantum computer 
\cite{cirac-zoller95,wineland98b,sorensen99,kielpinski02}.  Ground-state hyperfine levels within trapped
ions can act as nearly ideal quantum bit memories and be measured with essentially perfect quantum 
efficiency \cite{blatt88}.  Qubits based on trapped ions can be entangled by applying appropriate radiation that couples 
the internal levels of the ions with their collective quantum motion \cite{cirac-zoller95,monroe95b,molmer99,sackett00}.  
Such quantum logic gates are best realized with stimulated Raman transitions (SRT), involving two phase-coherent optical fields with frequency difference equal to the hyperfine splitting of the ion \cite{thomas82,kasevich92,monroe95a}.  These fields are far detuned from the excited state, making decoherence due to {\it spontaneous} Raman scattering negligible, though the SRT coupling itself vanishes when 
the detuning becomes much larger than the excited state fine-structure splitting\cite{wineland98b}.  We therefore desire qubit ions such as $^{111}Cd^+$ or $^{199}Hg^+$ that have a large fine-structure splitting.  Such ions
also exhibit a large ground-state hyperfine splitting, making it difficult to span the frequency difference 
with conventional acousto-optic modulators.  In this letter, we describe several methods for driving SRT in 
trapped $^{111}Cd^+$ ions using a high frequency electro-optic phase modulator (EOM). 

\begin{figure}
\includegraphics[width=8.4cm]{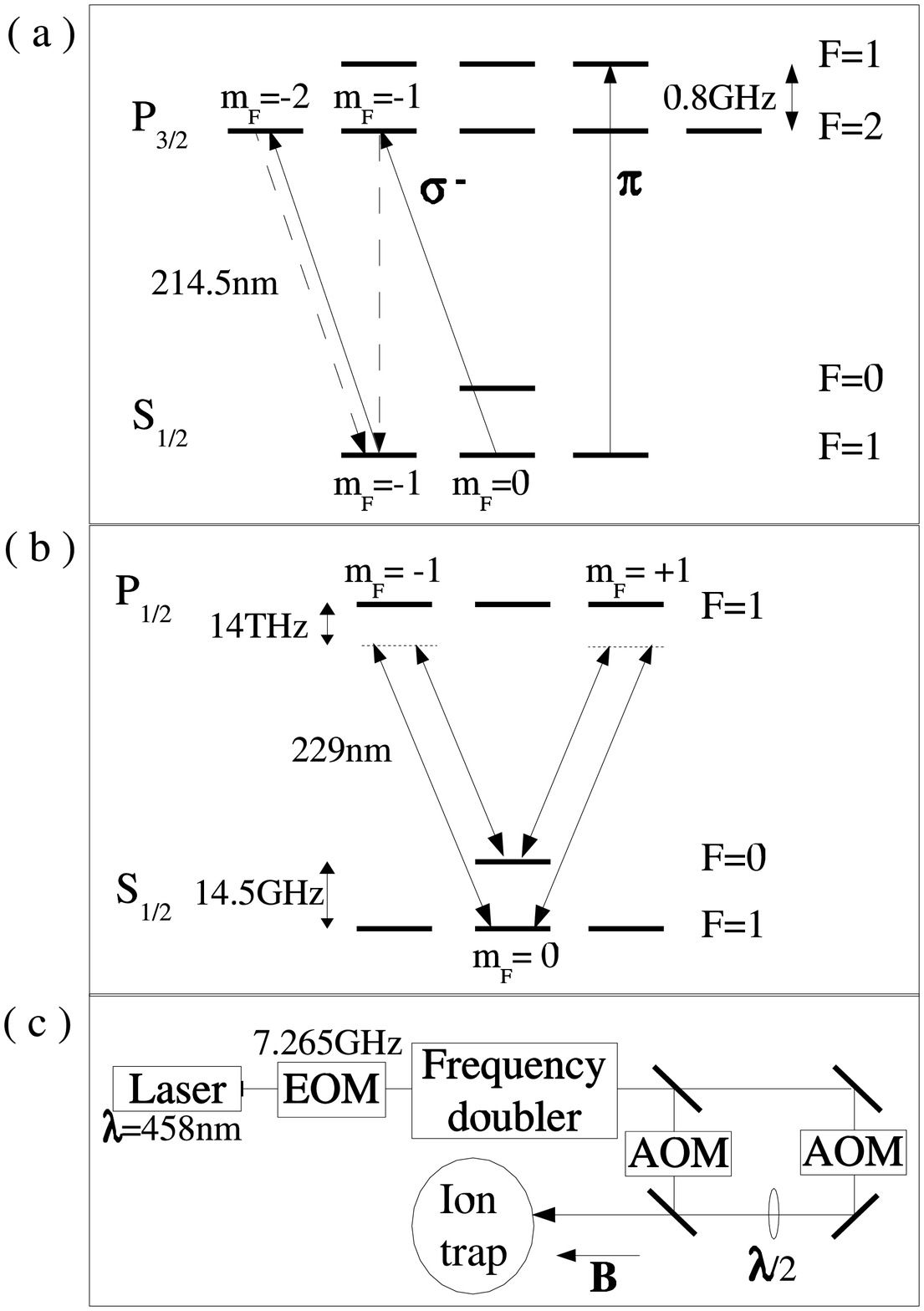}

\caption{(a) Relevant energy levels of a $^{111}Cd^+$ ion, with resonant detection($\sigma^-$-polarized) and optical pumping ($\pi$-polarized) transitions indicated.  (b) Coupling scheme for driving stimulated Raman transitions (SRT) simultaneously with both $\Delta m_F=\pm 1$ couplings.  (c) Schematic of the optical source for driving SRT, including Mach-Zehnder interferometer and AOM shifters to create frequency shift $\Delta \omega$ discussed in the text.  The $\lambda/2$-plate and the external magnetic field provide a constructive interference between the two paths in (b).  Alternatively, the beams could be circularly polarized, and only one of the paths in (b) is utilized. }
\label{setup}
\end{figure}

The experiment is conducted in an asymmetric quadrupole rf ion trap, as described in previous work \cite{blinov02}.  
Qubits are stored in the $^2S_{1/2}$ $|F=1,m_F=0\rangle$ and $|F=0,m_F=0\rangle$ ground state hyperfine levels 
of a single trapped $^{111}Cd^+$ ion (nuclear spin $I=1/2$), with a frequency splitting of 
$\omega_{HF} = 14.53\ GHz$\cite{tanaka96}, as shown in figure \ref{setup}b.  

\begin{figure}
\includegraphics[height=8.4cm,angle=270]{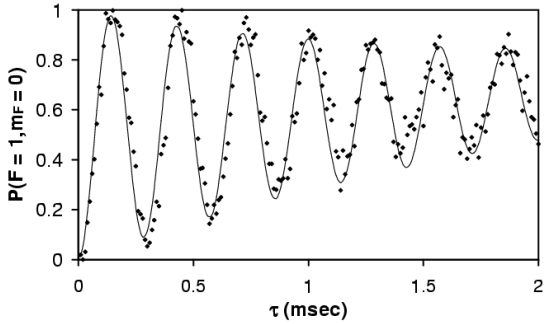}

\caption{Rabi flopping on a single ion using SRT with a 14THz detuning.  We prepare the ion in the $^2S_{1/2}$ $|F=0,m_F=0\rangle$ state and then pulse the Raman beams.  The probability of the ion in the $^2S_{1/2}$ $|F=1,m_F=0\rangle$ state (average number of photons collected) vs. Raman pulse time is plotted here, with the Rabi frequency being $\Omega\approx 2\pi \times 2 kHz$.  The decaying envelope is consistent with laser amplitude noise, and the slight upward trend is due to residual light leakage from the resonant measuring beam, causing optical pumping.  Each point represents an average of 100 experiments.}
\label{Rabi}
\end{figure}

Rabi oscillations are measured by performing the following sequence: (see Fig \ref{setup})(i) The ion is optically pumped to the 
$|F=0,m_F=0\rangle$ qubit state by $\pi$-polarized radiation 
resonant with the $^2S_{1/2} (F=1) \rightarrow \, ^2P_{3/2} (F=1)$ transition. 
(ii) SRT are driven
by applying the electro-optic modulated Raman beams to the ion for time $\tau$.
(iii) The qubit state is measured by collecting the ion fluorescence
from a $1\ ms$ pulse of $\sigma^-$-polarized radiation resonant with the cycling transition
$^2S_{1/2} (F=1) \rightarrow \, ^2P_{3/2} (F=2)$.  This sequence is 
repeated multiple times for each $\tau$, and the Rabi frequency $\Omega$ is extracted from the averaged fluorescence oscillation
in time (an example is shown in Figure \ref{Rabi}).  

We use a tunable frequency-quadrupled Ti:Sapphire laser operating near 214.5nm for the resonant optical 
pumping and detection steps (i) and (iii).  For the SRT in step (ii), a 458nm $Nd:YVO_4$ laser is phase-modulated with a resonant EOM at $\omega_{HF}/2\simeq 2\pi\times 7.265\ GHz$ and subsequently frequency doubled.  The blue optical field following the EOM can be written as
\begin{equation}
E_1=\frac{E_{0}}{2}e^{i(kx-\omega t)}\sum _{n=-\infty }^{\infty }J_{n}(\phi )e^{in((\delta k)x-\omega _{HF}t/2)} + c.c. ,\label{eq1}\end{equation}
where $E_0$ is the unmodulated field amplitude, $J_{n}(\phi)$ is the n-th order Bessel function with modulation index $\phi$, and $\delta k=\omega _{HF}/2c$. We couple the beam into a build-up cavity containing
a BBO crystal for sum frequency generation. The free spectral range (fsr) of the cavity is carefully tuned to be 1/4 of the modulation frequency so that the carrier and all the sidebands
resonate simultaneously.  The resulting ultraviolet radiation consists of a comb of frequencies centered at 229nm and separated by $\omega_{HF}/2$.  The carrier is detuned by $\Delta / 2 \pi \sim 14 THz$ from the excited $^2P_{1/2}$ state (the fine-structure splitting between $^2P_{1/2}$ and 
$^2P_{3/2}$ is $\sim 72\ THz$), resulting in an expected probability of spontaneous emission per Rabi cycle of $\sim 10^{-5}$.  At the output of the cavity the electric
field becomes
\begin{equation}
E_{2}=\eta \frac{E_{0}^{2}}{4}e^{2i(kx-\omega t)}\sum _{n=-\infty }^{\infty }J_{n}(2\phi )e^{in((\delta k)x-\omega _{HF}t/2)}+ c.c.,\label{eq2}\end{equation}
where $\eta$ is the harmonic conversion efficiency (assumed constant over all frequencies considered).  All pairs of spectral components of electric field separated by frequency $\omega_{HF}$ can individually drive SRT in the ion, but the net Rabi frequency vanishes due to a destructive interference.  Therefore, the relative phases and/or amplitudes of the spectral components in Eq \ref{eq2} must be modified in order to drive SRT.  We present three schemes below.

One approach is to employ a Mach-Zehnder (MZ) interferometer, where the beam is split and recombined at another location with path length difference $\Delta x$ (see figure \ref{setup}c). The expression for the Rabi frequency is
\begin{equation}
 \Omega =\frac{\mu _{1}\mu _{2}\left\langle E_2 E_2^{*}e^{i\omega _{HF}t}\right\rangle }{\hbar^2 \Delta }=\Omega_0e^{i(\delta k)(2x+\Delta x)} \sum _{n=-\infty }^{\infty }J_{n}(2\phi )J_{n-2}(2\phi )cos((2k+(n-1)\delta k)\Delta x),\label{eq:4}\end{equation}
where $\mu_{1}$ and $\mu_{2}$ are the matrix elements of the electric dipole moment for a transition between the respective hyperfine states and the excited state, and the fields are time-averaged under the rotating wave approximation ($\Omega << \omega_{HF}<<\Delta$).  The base Rabi frequency $\Omega_0=\frac{\mu _{1}\mu _{2}}{\hbar^2\Delta } \left| \eta E_{0}^{2}/4\right|^{2}$ pertains to the usual case of SRT with a pair of monochromatic Raman beams separated in frequency by $\omega_{HF}$ and each with field amplitude $\eta E_0^2/4$. For $(\delta k)\cdot \Delta x = (2j+1)\pi$, where $j$ is an integer, the Rabi frequency can be as high as $\Omega=0.487\Omega_0$ for $\phi=0.764$.  One drawback to the form of Eq \ref{eq:4} is that the $k \Delta x$ factor in the cosine requires optical stability of the MZ interferometer.  This can be circumvented by introducing a relative frequency shift $\Delta \omega >> \Omega$ between the two paths of the MZ.  This shift can be compensated by changing the modulation frequency of the EOM by $\pm \Delta \omega/2$, resulting in a Rabi frequency of
\begin{equation}
\Omega =\Omega_0 e^{-ik\Delta x}e^{-2i(\delta k) \Delta x}\sum _{n=-\infty }^{\infty }J_{n}(2\phi )J_{n-2}(2\phi )e^{in(\delta k)\Delta x},\label{eq:5}\end{equation}
where $\Delta k=\Delta \omega/2c << \delta k$.  Note that the cosine term has been replaced by a phase factor $e^{-ik\Delta x}$, thus eliminating the effects of small changes in $\Delta x$ on the magnitude of the Rabi frequency.  In this case, the SRT Rabi frequency can be as high as $\Omega = 0.244 \Omega_0$ for $\phi=0.764$ and $\delta k \Delta x = (2j+1)\pi$, where $j$ is an integer.

\begin{figure}
\includegraphics[height=8.4cm,angle=270]{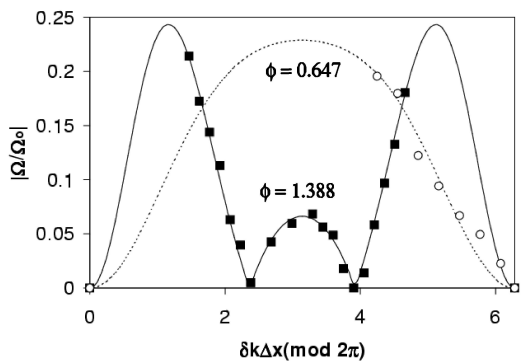}

\caption{Rabi frequency vs. Mach-Zehnder path length difference for two values of EOM modulation index $\phi$.  The lines are theory, and the data are fitted to Eq \ref{eq:5} using the y-axis scale and modulation index as parameters.  The fits agree with independent measurements of the modulation index using a Fabry-Perot cavity.}
\label{MZ}
\end{figure}

In the experiment, we set $\Delta \omega $ to be $2\pi \times 4 MHz$ and measure the Rabi frequency $\Omega$ as the relative path length $\Delta x$ of the MZ is varied.   We fit $\Omega$ to Eq \ref{eq:5} to extract the modulation index $\phi$, which is also independently measured with a Fabry-Perot spectrum analyzer.  The results	are plotted in Figure \ref{MZ}.  The dependence on gross path length difference with spatial period $\Delta x = 2 \pi / \delta k = 4.13 cm$ is clearly visible \cite{Chu92APB}. 

\begin{figure}
\includegraphics[width=8.4cm]{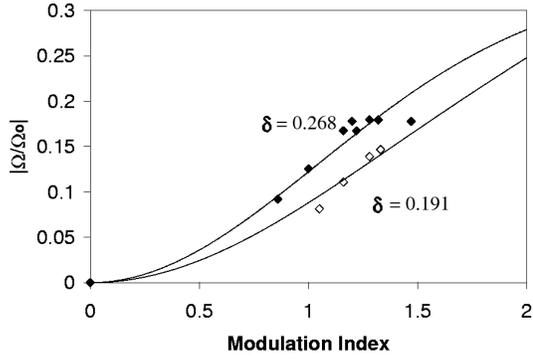}

\caption{Rabi frequency vs. EOM modulation index for two different detunings $\delta$ of the BBO build-up cavity free spectral range with respect to the EOM modulation frequency, scaled to the cavity linewidth (see text).  The lines are theory, and the data are fitted to Eq \ref{eq:7} using the y-axis scale and detuning as parameters.  The fits agree with independent measurements of the cavity free spectral range and EOM modulation frequency.}
\label{detune}
\end{figure}

We also shaped the sideband spectrum without using a MZ interferometer by detuning the fsr of the frequency doubling cavity slightly from a subharmonic of the EOM frequency.  This detuning modifies the amplitude and phase of each sideband with respect to the carrier, resulting in a Rabi frequency of
\begin{equation}
\Omega = 2 \Omega_0 e^{2i(\delta k)x}\sum _{n=-\infty }^{\infty }\sum _{m=-\infty }^{\infty }\sum _{l=-\infty }^{\infty }\left(\frac{J_{n-m}(\phi )}{1-i2(n-m)\delta}\right)\left(\frac{J_{m}(\phi )}{1-i2m\delta}\right)\left(\frac{J_{n+2-l}(\phi )}{1+i2(n+2-l)\delta}\right)\left(\frac{J_{l}(\phi )}{1+i2l\delta}\right),\label{eq:7}\end{equation}
where $\delta < 1$ is the number of cavity linewidths by which the first sideband is detuned from a cavity resonance.  Figure \ref{detune} displays $\Omega$ versus $\phi$ for various cavity detunings, and the data agree with Eq \ref{eq:7}.

Another possible scheme involves the suppression of certain spectral components.  By setting the fsr of the cavity to be $\omega _{HF}/(2n+1)$ where n is an integer, only alternating sidebands will resonate.  When the even or odd sidebands are selected, we find
\begin{equation}
\Omega_{even} =\Omega_0 e^{-2i(\delta k) x}\sum _{n=-\infty }^{\infty }\sum _{m=-\infty }^{\infty }\sum _{l=-\infty }^{\infty }J_{2(n-m)}(\phi )J_{2m}(\phi )J_{2(n+1-l)}(\phi )J_{2l}(\phi ).\label{eq:9}\end{equation}
\begin{equation}
\Omega_{odd} =\Omega_0 e^{-2i(\delta k) x}\sum _{n=-\infty }^{\infty }\sum _{m=-\infty }^{\infty }\sum _{l=-\infty }^{\infty }J_{2(n-m)+1}(\phi )J_{2m+1}(\phi )J_{2(n+1-l)+1}(\phi )J_{2l+1}(\phi ).\label{eq:10}\end{equation}
The maximum $\Omega_{even}$ is $0.230\Omega_0$ at modulation frequency $\phi=1.173$, while the maximum $\Omega_{odd}$ is $0.279\Omega_0$ at $\phi=1.603$.

The techniques reported here have been extended to drive SRT between internal and motional states of trapped ions for entangling quantum logic operations.  Here, the modulated Raman beams are split with a beamsplitter and recombined at the ion with non-copropagating wavevectors.  The resulting MZ interferometer gives rise to Rabi oscillations according to Eqs \ref{eq:4} or \ref{eq:5}, with the additional dependence on the motional state of the ion\cite{wineland98b}.  In addition, by inserting an appropriate relative frequency shift $\Delta \omega$ between the arms of the interferometer, the resulting pair of offset frequency combs can be used to implement the Molmer-Sorensen bichromatic quantum logic gate scheme\cite{molmer99, sorensen99}.

In summary, we have used a high-frequency EOM to demonstrate SRT Rabi oscillations on a single trapped $^{111}Cd^+$ ion with a frequency separation of $14.53\ GHz$, where spontaneous emission is expected to be negligible.  The techniques of shaping the sideband spectrum or creating an interference pattern between multiple modulated beams results in efficient use of the optical power for SRT.  This new tool for coherent control of single ions brings us one step forward in buiding higher fidelity quantum logic gates.

We wish to acknowledge useful discussions with Ralph Conti and Chitra Rangan.  This work was supported by the U.S. National Security Agency and the Advanced Research and Development Activity under U.S. Army Research Office contract DAAD19-01-1-0667, and the National Science Foundation ITR Program.

\newpage

\end{document}